\documentclass[12pt,journal,onecolumn]{IEEEtran}
\IEEEoverridecommandlockouts
\usepackage[T1]{fontenc} 
\usepackage{times}
\usepackage[cmex10]{amsmath}
\usepackage{amssymb}
\usepackage{amsthm}
\usepackage{color}
\usepackage{algorithm}

\usepackage{graphicx}
\usepackage[caption=false]{subfig}
\usepackage{multirow}
\usepackage{algpseudocode}

\usepackage[bookmarks=false,colorlinks=false,pdfborder={0 0 0}]{hyperref}
\usepackage{cite}
\usepackage{bm}
\usepackage{arydshln}
\usepackage{mathtools}
\usepackage{microtype}

\usepackage[figuresright]{rotating}
\usepackage{threeparttable}
\usepackage{booktabs}
\usepackage{verbatim}

\newcommand{\overbar}[1]{\,\overline{\!{#1}}}

\newtheorem{theorem}{Theorem}

\newtheorem{lemma}[theorem]{Lemma}

\def\BibTeX{{\rm B\kern-.05em{\sc i\kern-.025em b}\kern-.08em
		T\kern-.1667em\lower.7ex\hbox{E}\kern-.125emX}}

\long\def\symbolfootnote[#1]#2{\begingroup
	\def\thefootnote{\fnsymbol{footnote}}\footnote[#1]{#2}\endgroup}
\IEEEoverridecommandlockouts

\title{Reed-Solomon Codes over Cyclic Polynomial Ring with Lower Encoding/Decoding Complexity}

\author{Wenhao Liu$^\S$, Zhengyi Jiang$^\S$, Zhongyi Huang$^\S$\, Linqi Song$^{\star}$, Hanxu Hou$^\dagger$$^\ddagger$\\
$^\S$ Department of Mathematical Sciences, Tsinghua University\\
 $^{\star}$ Department of Computer Science, City University of Hong Kong \\
$^\dagger$ Dongguan University of Technology
}

\begin{document}
\let\emph\textit
\maketitle
\pagestyle{empty}  
\thispagestyle{empty} 
\begin{abstract}\symbolfootnote[0]{$^\ddagger$:  Corresponding author. This work was partially supported by the Key-Area Research and Development Program of Guangdong Province 2020B0101110003, National Key R\&D Program of
China (No. 2020YFA0712300), the National Natural Science Foundation of China (No. 62071121, 62371411, 12025104),
Basic Research Enhancement Program of China under Grant 2021-JCJQ-JJ-0483.}Reed-Solomon (RS) codes are constructed over a finite field that have been
widely employed in storage and communication systems.
Many fast encoding/decoding algorithms such as fast Fourier transform (FFT) and modular approach are designed for RS codes to reduce the encoding/decoding complexity defined as the number of XORs involved in the encoding/decoding procedure.
In this paper, we present the construction of RS codes over the cyclic polynomial ring $ \mathbb{F}_2[x]/(1+x+\ldots+x^{p-1})$ and show that our codes are maximum distance separable (MDS) codes.
Moreover, we propose the FFT and modular approach over the ring that can be employed in our codes for encoding/decoding complexity reduction. We show that our codes have 17.9\% encoding complexity reduction and 7.5\% decoding complexity reduction compared with RS codes over finite field, for $(n,k)=(2048,1984)$.

\end{abstract}


\IEEEpeerreviewmaketitle

\section{Introduction}
\label{sec:intro}
{\em Reed-Solomon} (RS) codes are a class of maximum distance separable (MDS) codes that have been widely employed in communication and storage systems \cite{1960ReedPolynomial}.
An $(n,k)$ RS code encodes $k$ data symbols to obtain the $n$ codeword symbols over the field $\mathbb{F}_{2^q}$, where $n,k,q$ are positive integers, $n>k$ and $n\leq 2^q$. In general, we can construct RS codes by choosing the Vandermonde matrix as generator matrix and the codes are MDS if and only if $n\leq 2^q$. 

Encoding/decoding complexity defined as the total number of XORs involved in the encoding/decoding (error correction) procedure is the key metric for RS codes. Field addition and field multiplication are two basic operations in the encoding/decoding procedure, where one field
addition over $\mathbb{F}_{2^q}$ requires $q$ XORs and one field multiplication over $\mathbb{F}_{2^q}$ requires much more XORs.
Many fast encoding/decoding algorithms \cite{2015FasterAlgorithms,2020FastEncoding, 2022NewEncoding, 2023EfficientInterpolation,1993NewArrayCodes,hou2016,hou2018a,Hou2018form,hou2021generalization} have been proposed for encoding/decoding complexity reduction. We can divide the existing fast encoding/decoding algorithms into two methods. One method is to reduce the number of basic field operations, including addition and multiplication operations, such as \cite{2015FasterAlgorithms,2020FastEncoding, 2022NewEncoding, 2023EfficientInterpolation}. To the best of our knowledge, the fast Fourier transform (FFT) and modular approach \cite{2022NewEncoding} achieve the best asymptotic number of field operations, where the number of field operations involved in encoding procedure is $O(n\log(n-k))$  and the number of field operations involved in decoding procedure is $O(n\log(n-k)+(n-k)\log^2(n-k))$. 
Another method \cite{1993NewArrayCodes,hou2016,hou2018a,Hou2018form,hou2021generalization} is to construct RS codes over the cyclic polynomial ring $ \mathbb{F}_2[x]/(1+x+\ldots+x^{p-1})$ or more general ring that can avoid the expensive field multiplication.
We will present new fast encoding/decoding algorithms for RS codes by jointly considering the above two methods.


The main contributions of the paper are summarized as follows. First, we propose RS codes over the cyclic polynomial ring $\mathbb{F}_2[x]/(1+x+\ldots +x^{p-1})$ and show that our codes are MDS codes. Second, we propose fast encoding/decoding algorithms for our codes by employing the FFT and modular approach over the ring $\mathbb{F}_2[x]/(1+x+\ldots +x^{p-1})$ in the encoding/decoding procedure. We show that our codes can achieve the best asymptotic number of field operations required in the encoding/decoding procedure. Third, we evaluate the encoding/decoding complexity for our codes and the RS codes over finite field to show that our codes have lower encoding/decoding complexity.
Specifically, our codes can reduce $17.9\%$ encoding complexity and $7.5\%$ decoding complexity compared with RS codes over the field $\mathbb{F}_{2^{11}}$ with the fast encoding/decoding algorithms in \cite{2022NewEncoding}, for $(n,k)=(2048, 1984)$. 


\section{RS Codes over Cyclic Polynomial Ring}
\label{sec:Upperbound}
In this section, we present the construction of our RS codes over the cyclic polynomial ring $\mathbb{F}_2[x]/(1+x+\ldots+x^{p-1})$, where $p$ is an odd number. Then we show our codes are MDS codes.
Throughout the paper, we use the notation $M_p(x)=1+x+\ldots+x^{p-1}$ and $[t]=\{1,2,\ldots,t\}$ for positive integer $t$.
Let $|S|$ be the number of elements in set $S$.

\subsection{Construction}\label{subsec:Construction}
We denote the ring $\mathbb{F}_2[x]/(M_p(x))$ by $\mathcal{R}_p$ and our codes are operated over $\mathcal{R}_p$, where $p$ is an odd number. We can factorize $M_p(x)$ as a product of $t$ distinct irreducible polynomials over $\mathbb{F}_2$, i.e., $$M_p(x)=\prod\limits_{i=1}^t p_i(x),$$
where $t\geq 1$ and $p_i(x)\in \mathbb{F}_2[x]$ for $i\in[t]$. 

By the Chinese remainder theorem, the ring $\mathcal{R}_p$ is isomorphic to the direct sum of the $t$ fields $\mathbb{F}_2[x]/(p_1(x))$, $\mathbb{F}_2[x]/(p_2(x))$, $\ldots$, $\mathbb{F}_2[x]/(p_t(x))$ \cite[Theorem 6]{HOU2019}. The isomorphism $\Phi:\mathcal{R}_p\rightarrow$\\$(\mathbb{F}_2[x]/(p_1(x)),\ldots,\mathbb{F}_2[x]/(p_t(x)))$ is defined by 
\begin{equation}\label{eq:ChineseRemainder}
    \begin{aligned}
    \Phi(r(x))=
 (r(x)\bmod p_1(x),\ldots,r(x)\bmod p_t(x)),
    \end{aligned}
\end{equation}
where $r(x)\in \mathcal{R}_p$. 
For element $r\in \mathcal{R}_p$, denote the $i$-th component of $\Phi(r)$ as $r^{(i)}$, i.e., $\Phi(r)=(r^{(1)},r^{(2)},\ldots,r^{(t)})$.
We can directly obtain the following result without proof.
\begin{lemma}\label{lemma:invertible}
The polynomial $r(x)\in \mathcal{R}_p$ is invertible if and only if $\forall i \in [t],r(x)\not\equiv0\pmod{p_i(x)}$.
\end{lemma}
    

Next, we provide the definition of our $(n=2^m, k=2^m-2^\mu)$ RS codes over $\mathcal{R}_p$, where \(m=\gcd(\deg(p_1(x)),\ldots,\deg(p_t(x)))\) and $\mu\in \{0,1,\ldots,m-1\}$.

For $i=1,2,\ldots,t$, since $\lvert\mathbb{F}_2[x]/(p_i(x))\rvert=2^{\deg(p_i)}$ and $m\mid \deg(p_i)$,  there exists a unique subfield of size $2^m$ in $\mathbb{F}_2[x]/(p_i(x))$. We choose a basis of the subfield and denote it as $\{v_j^{(i)}\}_{j=0,1,\ldots,m-1}$. For $j=0,1,\ldots,m-1$, let $v_j=\Phi^{-1}(v_j^{(1)},v_j^{(2)},\ldots,v_j^{(t)})$, where $\Phi^{-1}$ is the inverse mapping of $\Phi$ which is in Eq. \eqref{eq:ChineseRemainder}. For \(l=0,1,\ldots,2^m-1\), define \begin{equation}\label{eq:LCHbasis}
    \omega_{l}=l_0v_0+l_1v_1+\ldots+l_{m-1}v_{m-1},
\end{equation} where $l_0,l_1,\ldots,l_{m-1}\in \{0,1\}$ and $(l_0,l_1,\ldots,l_{m-1})$ is the binary representation of $l$.


Given the $k$ data symbols $g_0,g_1,\ldots,g_{k-1}\in\mathcal{R}_p$, we will proof in Theorem \ref{theorem:codeLength} that there exists a polynomial $f(x)=\sum_{i=0}^{k-1}f_ix^i\in \mathcal{R}_p[x]$ such that $f(\omega_{2^\mu+i})=g_i$ for $i=0,1,\ldots,k-1$ and we define the $n$ codeword symbols as
\[
    (f(\omega_0),f(\omega_1),\ldots,f(\omega_{n-1})),
\]
thus our codes are systematic codes. The codes defined above can be viewed as replacing the data symbols with $\{f_i\}_{i=0}^{k-1}$ and constructing codes by the generator matrix
\begin{small}
\begin{equation}\label{eq:GeneratorMatrix}
    \begin{pmatrix}
        1&1&\cdots&1\\
        \omega_0&\omega_1&\cdots&\omega_{n-1}\\
        \vdots&\vdots&\ddots&\vdots\\
        \omega_0^{k-1}&\omega_1^{k-1}&\cdots&\omega_{n-1}^{k-1}
    \end{pmatrix}.
\end{equation}    
\end{small}

For $l=0,1,\ldots,n-1$, denote $\Phi(\omega_l)=(\omega_l^{(1)},\omega_l^{(2)},\ldots,\omega_l^{(t)})$. We can compute that \(\omega_l^{(i)}=l_0v_0^{(i)}+l_1v_1^{(i)}+\ldots+l_{m-1}v_{m-1}^{(i)}\), where $i=1,2,\ldots,t$ and $(l_0,l_1,\ldots,l_{m-1})$ is the binary representation of $l$.
The next theorem shows that our codes are well-defined and are MDS codes.

\begin{theorem}\label{theorem:codeLength}
Our codes are well-defined and are MDS codes.
\end{theorem}

\begin{IEEEproof}
Firstly, we proof that the determinant of any 
$k\times k$ submatrix of the generator matrix in Eq. \eqref{eq:GeneratorMatrix} is invertible in $\mathcal{R}_p$. This is equivalent to show that the determinant $$\prod\limits_{0\le l<t\le k-1}(\omega_{j_l}-\omega_{j_t})$$ 
is invertible in $\mathcal{R}_p$, where $0\le j_0\le j_1\le \cdots \le j_{k-1}\le n-1$. And it is equivalent to proof that $\forall\  0\ \le i<j\le {n-1}$, $\omega_{i}-\omega_{j}$ is invertible in $\mathcal{R}_p$. For any $j\neq k$, we can show
\[
\omega_j-\omega_k\equiv \omega_j^{(i)}-\omega_k^{(i)}\not\equiv0\pmod{p_{i}(x)},\forall i\in[t],
\] by the fact that the binary representations of $j$ and $k$ are different and \(v_0^{(i)},v_1^{(i)},\ldots,v_{m-1}^{(i)}\) are linearly independent for any $i\in[t]$. Therefore, $\omega_j-\omega_k$ is invertible by Lemma \ref{lemma:invertible}, and the determinant of any 
$k\times k$ submatrix of the generator matrix in Eq. \eqref{eq:GeneratorMatrix} is invertible in $\mathcal{R}_p$.

Given the $k$ data symbols $g_0,g_1,\ldots,g_{k-1}\in\mathcal{R}_p$, finding a polynomial $f(x)=\sum_{i=0}^{k-1}f_ix^i\in \mathcal{R}_p[x]$ such that $f(\omega_{2^\mu+i})=g_i$ for $i=0,1,\ldots,k-1$ is equivalent to solving the equation
\begin{equation}
    \begin{pmatrix}
        f_0\\
        f_1\\
        \vdots\\
        f_{k-1}
    \end{pmatrix}^T\begin{pmatrix}
        1&1&\cdots&1\\
        \omega_{n-k}&\omega_{n-k+1}&\cdots&\omega_{n-1}\\
        \vdots&\vdots&\ddots&\vdots\\
        \omega_{n-k}^{k-1}&\omega_{n-k+1}^{k-1}&\cdots&\omega_{n-1}^{k-1}
    \end{pmatrix}=\begin{pmatrix}
        g_0\\
        g_1\\
        \vdots\\
        g_{k-1}
    \end{pmatrix}.
\end{equation} Since the determinant of any 
$k\times k$ submatrix of the generator matrix in Eq. \eqref{eq:GeneratorMatrix} is invertible, the polynomial $f(x)$ exists and is unique, thus our codes are well-defined.

Similar to the proof in \cite{1993NewArrayCodes}, our codes are MDS codes if and only if the determinant of any 
$k\times k$ submatrix of the generator matrix in Eq. \eqref{eq:GeneratorMatrix} is invertible in $\mathcal{R}_p$. Therefore, our codes are MDS codes.

\end{IEEEproof}

\section{Encoding and Decoding Methods}
\label{sec:encoding}

In this section, we present FFT and the inverse FFT (IFFT) algorithms over $\mathcal{R}_p$. Then the encoding/decoding algorithms based on FFT and modular approach for our codes are presented. By a small abuse of the notation, we use the same indeterminate $\alpha$ when referring to  polynomials as elements in $\mathcal{R}_p$ or $\mathbb{F}_2[x]/(p_i(x)),\forall i \in [t]$. When the modulo operation is applied to a vector, it represents performing the modulo operation on each component.

For simplicity, we assume $m=\deg(p_1(x))=\deg(p_2(x))=\cdots=\deg(p_t(x))$ in this section, in which case the subfield generated by $\{\omega_l^{(i)}\}_{l=0,1,\ldots,n-1}$ is identical to $\mathbb{F}_2[x]/(p_i(x))$ for $i=1,2,\ldots,t$. 
Otherwise, the same argument holds by replacing $\mathbb{F}_2[x]/(p_i(x))$ with the subfield generated by $\{\omega_l^{(i)}\}_{l=0,1,\ldots,n-1}$ for $i=1,2,\ldots,t$.

\subsection{FFT and IFFT Algorithms over $\{\mathbb{F}_2[x]/(p_i(x))\}_{i\in[t]}$ and $\mathcal{R}_p$}

The FFT and IFFT algorithms over $\{\mathbb{F}_2[x]/(p_i(x))\}_{i\in[t]}$ and $\mathcal{R}_p$ are presented as follows.

\begin{enumerate}
    \item[(1)][FFT and IFFT Algorithms over $\{\mathbb{F}_2[x]/(p_i(x))\}_{i\in[t]}$]

For $i=1,2,\ldots,t$, the subspace polynomial \cite{2010SubspacePoly} over $\mathbb{F}_2[x]/(p_i(x))$ is defined by $s_\tau^{(i)}(x)=\prod\limits_{l=0}^{2^\tau-1}(x-\omega_l^{(i)})$ for $\tau=0,1,\ldots,m$. For $l=0,1,\ldots,2^m-1$, define \[\overbar{X}_l^{(i)}(x)=\frac{s_0^{(i)}(x)^{l_0}s_1^{(i)}(x)^{l_1}\cdots s^{(i)}_{m-1}(x)^{l_{m-1}}}{s_0^{(i)}(v_0^{(i)})^{l_0}s_1^{(i)}(v_1^{(i)})^{l_1}\cdots s^{(i)}_{m-1}(v_{m-1}^{(i)})^{l_{m-1}}},\] where $(l_0,l_1,\ldots,l_{m-1})$ is the binary representation of $l$. 
To avoid confusion, we use $\alpha$ instead of $x$ to denote $\mathbb{F}_2[x]/(p_i(x))$ when it comes to polynomial ring $(\mathbb{F}_2[\alpha]/(p_i(\alpha)))[x]$. Since the degree of $\overbar{X}_l^{(i)}(x)$ is $l$ for $l=0,1,\ldots,2^m-1$, the set $\overbar{X}^{(i)}=\{\overbar{X}_0^{(i)},\overbar{X}_1^{(i)},\ldots,\overbar{X}_{2^m-1}^{(i)}\}$ is a basis of $(\mathbb{F}_2[\alpha]/(p_i(\alpha)))[x]/(x^{2^m}-x)$. For any $\tau\in\{0,1,\ldots,m\}$, any $f(x)\in(\mathbb{F}_2[\alpha]/(p_i(\alpha)))[x]$ with \(\deg(f(x)) < 2^\tau\) can be uniquely represented as the linear combination $f(x)=\sum\limits_{l=0}^{2^\tau-1}f_l\overbar{X}_l^{(i)}(x)$. The vector $\overbar{f}=(f_0,f_1,\ldots,f_{2^\tau-1})$ is the coordinate vector of $f(x)$ with respect to the basis $\overbar{X}^{(i)}$. Then the FFT over the field $\mathbb{F}_2[x]/(p_i(x))$, denoted by $\text{FFT}_{\overbar{X}^{(i)}}$, is defined by
\begin{align*}
    \text{FFT}_{\overbar{X}^{(i)}}(\overbar{f},\tau,\beta)
    =(f(\omega_0^{(i)}+\beta),f(\omega_1^{(i)}+\beta),\ldots,f(\omega_{2^\tau-1}^{(i)}+\beta)),
\end{align*} where $\beta\in \mathbb{F}_2[x]/(p_i(x))$. 

Since the set $\{\omega_0^{(i)},\omega_1^{(i)},\ldots,\omega_{2^m-1}^{(i)}\}$ forms a finite field of size $2^m$ and \(\omega_l^{(i)}=l_0v_0^{(i)}+l_1v_1^{(i)}+\ldots+l_{m-1}v_{m-1}^{(i)}\) holds, $\text{FFT}_{\overbar{X}^{(i)}}$ is well-defined and can be calculated by \cite[Algorithm 2]{2022NewEncoding}.

    \item[(2)][FFT and IFFT Algorithms over $\mathcal{R}_p$]
    
    Similarly, define $s_\tau(x)=\prod\limits_{l=0}^{2^\tau-1}(x-\omega_l)\in\mathcal{R}_p[x]$ for $\tau=0,1,\ldots,m$. For $l=0,1,\ldots,2^m-1$, define\[\overbar{X}_l(x)=\frac{s_0(x)^{l_0}s_1(x)^{l_1}\cdots s_{m-1}(x)^{l_{m-1}}}{s_0(v_0)^{l_0}s_1(v_1)^{l_1}\cdots s_{m-1}(v_{m-1})^{l_{m-1}}},\] where $(l_0,l_1,\ldots,l_{m-1})$ is the binary representation of $l$. Since we proved that for any $i\ne j$, $\omega_i-\omega_j$ is invertible in $\mathcal{R}_p$ in Theorem \ref{theorem:codeLength}, we can show that $s_i(v_i)$ is invertible in $\mathcal{R}_p$ for $i=0,1,\ldots,m-1$. Therefore, $\overbar{X}_l(x)$ is well defined, the degree of $\overbar{X}_l(x)$ is $l$ and the leading coefficient is invertible. Then the set $\overbar{X}=\{\overbar{X}_0,\overbar{X}_1,\ldots,\overbar{X}_{2^m-1}\}$ forms a basis of $\mathcal{R}_p[x]/(x^{2^m}-x)$. Similarly, for any $\tau\in\{0,1,\ldots,m\}$ and $f(x)\in \mathcal{R}_p[x]$ with \(\deg(f(x)) < 2^\tau\), the coordinate vector of $f(x)$ with respect to the basis $\overbar{X}$ is denoted as $\overbar{f}$. Then the FFT over $\mathcal{R}_p$, denoted by $\text{FFT}_{\overbar{X}}$, is defined by
\begin{align*}
    \text{FFT}_{\overbar{X}}(\overbar{f},\tau,\beta)
    =(f(\omega_0+\beta),f(\omega_1+\beta),\ldots,f(\omega_{2^\tau-1}+\beta)),
\end{align*} where $\beta\in \mathcal{R}_p$. 

Next we provide the algorithm for computing $\text{FFT}_{\overbar{X}}(\overbar{f},\tau,\beta)$ and prove its correctness. The FFT algorithm over $\mathcal{R}_p$ is presented in Algorithm \ref{alg:FFT}, where $f(x)=\sum\limits_{l=0}^{2^\tau-1}f_l\overbar{X}_l(x)\in \mathcal{R}_p[x]$ and $\deg(f(x))<2^\tau$. We proof in Theorem~\ref{theorem:FFT_alg_correct} that Algorithm \ref{alg:FFT} exactly output $\text{FFT}_{\overbar{X}}(\overbar{f},\tau,\beta)$.
\begin{small}
\begin{algorithm}
\caption{$\text{FFT}_{\overbar{X}}(\overbar{f},\tau,\beta)$ over $\mathcal{R}_p$}\label{alg:FFT}
\begin{algorithmic}[1]
\Require $\overbar{f}$ = (${f}_0$,${f}_1$,\ldots,${f}_{2^\tau-1}$), $\tau$, $\beta$
\Ensure $(f(\omega_0+\beta),f(\omega_1+\beta)),\ldots,f(\omega_{2^\tau-1}+\beta))$ 
\If {$\tau$ = 0}
\State \Return {${f}_0$} 
\EndIf
\For{$l$ = 0,1,\ldots,$2^{\tau-1}-1$}
\State $a_l^0={f}_l+\frac{s_{\tau-1}(\beta)}{s_{\tau-1}(v_{\tau-1})}{f}_{l+2^{\tau-1}}$
\State $a_l^1=a_l^0+{f}_{l+2^{\tau-1}}$
\EndFor
\State $a^0$ = $(a_0^0,a_1^0,\ldots,a_{2^{\tau-1}-1}^0)$, $a^1$ = $(a_0^1,a_1^1,\ldots,a_{2^{\tau-1}-1}^1)$
\State $A_0$ = $\text{FFT}_{\overbar{X}}(a^0, \tau-1,\beta)$, $A_1$ = $\text{FFT}_{\overbar{X}}(a^1, \tau-1,v_{\tau-1}+\beta)$
\State\Return $(A_0,A_1)$
\end{algorithmic}
\end{algorithm}
\end{small}

\begin{theorem}
    \label{theorem:FFT_alg_correct}
    Algorithm \ref{alg:FFT} exactly output $\text{FFT}_{\overbar{X}}(\overbar{f},\tau,\beta)=(f(\omega_0+\beta),f(\omega_1+\beta)),\ldots,f(\omega_{2^\tau-1}+\beta))$.
\end{theorem}

\begin{IEEEproof}
    Denote $\Phi(f_j)=(f_{j}^{(1)},f_{j}^{(1)},\ldots,f_{j}^{(t)})$ for $j=0,1,\ldots,2^\tau-1$ and $\Phi(\beta)=(\beta^{(1)},\beta^{(2)},\ldots,\beta^{(t)})$. For $i=1,2,\ldots,t$, define $f^{(i)}(x) = \sum\limits_{j=0}^{2^\tau-1}{f}_j^{(i)}\overbar{X}_j^{(i)}(x)$ and denote $\overbar{f^{(i)}}=
    ({f}_0^{(i)},{f}_1^{(i)},\ldots,{f}_{2^\tau-1}^{(i)})$ as the coefficient vector of $f^{(i)}(x)$ with respect to the basis $\overbar{X}^{(i)}$. 
    
    For $i=1,2,\ldots,t$, taking all elements in the computation process modulo $p_i(\alpha)$ component-wise, the input is $\overbar{f^{(i)}}=
    ({f}_0^{(i)},{f}_1^{(i)},\ldots,{f}_{2^\tau-1}^{(i)})$. Therefore, the input of Algorithm \ref{alg:FFT} is identical to the input of $\text{FFT}_{\overbar{X}^{(i)}}(\overbar{f^{(i)}},\tau,\beta^{(i)})$ in \cite[Algorithm 2]{2022NewEncoding} when taken modulo $p_i(\alpha)$.

    For $i=1,2,\ldots,t$, it is straightforward to verify that the intermediate steps of this algorithm are also identical to the $\text{FFT}_{\overbar{X}^{(i)}}(\overbar{f^{(i)}},\tau,\beta^{(i)})$ algorithm in \cite[Algorithm 2]{2022NewEncoding} under modulo $p_i(\alpha)$ by the following equation $$\frac{s_{\tau-1}(\beta)}{s_{\tau-1}(v_{\tau-1})}\equiv\frac{s_{\tau-1}^{(i)}(\beta^{(i)})}{s_{\tau-1}^{(i)}(v^{(i)}_{\tau-1})}\pmod{p_i(\alpha)}.$$ 

    Based on the above discussion, the output of Algorithm \ref{alg:FFT} is equivalent to the output of\\ $\text{FFT}_{\overbar{X}^{(i)}}(\overbar{f^{(i)}},\tau,\beta^{(i)})$ under modulo $p_i(\alpha)$ for $i=1,2,\ldots,t$. Assuming the output of Algorithm \ref{alg:FFT} is $(r_1,\ldots,r_{2^\tau-1})$, then we have
    \begin{equation}\label{eq:FFTproof}
        \begin{aligned}
            (r_1,\ldots,r_{2^\tau-1}) 
            \equiv (f^{(i)}(\omega_0^{(i)}+\beta^{(i)}),f^{(i)}(\omega_1^{(i)}+\beta^{(i)})),
            \ldots,f^{(i)}(\omega_{2^\tau-1}^{(i)}+\beta^{(i)})) \pmod{p_i(\alpha)}.
        \end{aligned}
        \end{equation}
        Since $\Phi$ is a ring isomorphism, it must be the case that $r_j = \Phi^{-1}(f^{(1)}(\omega_j^{(1)}+\beta^{(1)}),f^{(2)}(\omega_j^{(2)}+\beta^{(2)}),\ldots,f^{(t)}(\omega_j^{(t)}+\beta^{(t)}))=f(\omega_j+\beta)$ for $j=0,1,\ldots,2^\tau-1$. Therefore Algorithm \ref{alg:FFT} exactly output $(r_1,\ldots,r_{2^\tau-1})=(f(\omega_0+\beta),f(\omega_1+\beta),\ldots,f(\omega_{2^\tau-1}+\beta))$.

\end{IEEEproof}


\end{enumerate}

The IFFT algorithm over $\{\mathbb{F}_2[x]/(p_i(x))\}_{i\in[t]}$ and $\mathcal{R}_p$ can be defined similarly.


\subsection{Encoding Algorithm}


Given the $k$ data symbols $g_0,g_1,\ldots,g_{k-1}\in\mathcal{R}_p$, there exists $f(x)=\sum_{i=0}^{k-1}f_ix^i\in \mathcal{R}_p[x]$ such that $f(\omega_{2^\mu+i})=g_i$ for $i=0,1,\ldots,k-1$ by the construction in Section \ref{sec:Upperbound}. For $j=1,2,\ldots,2^{m-\mu}$, denote $$F_j=(f(\omega_{(j-1)\cdot 2^\mu}),f(\omega_{(j-1)\cdot 2^\mu+1}),\ldots,f(\omega_{j\cdot 2^\mu-1})).$$ By the definition of our codes, $F_1$ are the $2^{\mu}$ parity symbols and $F_2,F_3,\ldots,F_{2^{m-\mu}}$ are the $2^m-2^{\mu}$ data symbols. Then we only need to compute the parity symbols $F_1$.

For any $f(x)\in \mathcal{R}_p[x]$ with $\deg(f(x))=d$, i.e., $f(x)=\sum\limits_{j=0}^df_jx^j$, denote $f^{(i)}(x)\in (\mathbb{F}_2[\alpha]/(p_i(\alpha)))[x]$ by
\begin{equation}\label{eq:f^idefinition}
    f^{(i)}(x)=\sum\limits_{j=0}^d f_j^{(i)}x^j,
\end{equation}
where $i=1,2,\ldots,t$ and $\Phi(f_j)=(f_{j}^{(1)},f_{j}^{(2)},\ldots,f_{j}^{(t)})$ for $j=0,1,\ldots,d$.

For $i=1,2,\ldots,t$ and $j=1,2,\ldots,2^{m-\mu}$, define $F^{(i)}=(f^{(i)}(\omega_0^{(i)}), f^{(i)}(\omega_1^{(i)}), \ldots, f^{(i)}(\omega_{n-1}^{(i)}))$ and
    $$
        F_j^{(i)}=(f^{(i)}(\omega_{(j-1)\cdot 2^\mu}),f^{(i)}(\omega_{(j-1)\cdot 2^\mu+1}),\ldots,f^{(i)}(\omega_{j\cdot 2^\mu-1})).
    $$
For $i=1,2,\ldots,t$, $F^{(i)}$ is a codeword of RS codes over the field $\mathbb{F}_2[x]/(p_i(x))$ whose data symbols are $\{F_j^{(i)}\}_{j=2}^{2^{m-\mu}}$. The next Lemma shows the formula for calculating $F_1$.

\begin{lemma}\label{Lemma:lemma3}

The parity symbols of our RS codes can be calculated as follows,
        \begin{align*}\label{eq:lemma3}
        F_1=
        \text{FFT}_{\overbar{X}}(\text{IFFT}_{\overbar{X}}(F_2,\mu,\omega_{2^\mu})+\cdots
        +\text{IFFT}_{\overbar{X}}(F_{2^{m-\mu}},\mu,\omega_{2^{m}-2^\mu}),\mu,0).
    \end{align*}
\end{lemma}

\begin{IEEEproof}
    By the definition of $\text{FFT}_{\overbar{X}}$ and $\text{FFT}_{\overbar{X}^{(i)}}$, the following equation can be concluded for any $f\in \mathcal{R}_p[x]$ and $i\in[t]$,
    \begin{equation}\label{eq:FFTlemma}
        \text{FFT}_{\overbar{X}}(\overbar{f},\mu,\beta)\equiv \text{FFT}_{\overbar{X}^{(i)}}(\overbar{f^{(i)}},\mu,\beta^{(i)})\pmod{p_i(\alpha)},
    \end{equation} 
    where $\overbar{f^{(i)}}$ denotes the coordinate vector of $f^{(i)}(x)$ with respect to the basis $\overbar{X}^{(i)}$ and modulo is taken component-wise.
    
    Similarly, \begin{equation}\label{eq:IFFTlemma}
        \text{IFFT}_{\overbar{X}}(\overbar{f},\mu,\beta)\equiv \text{IFFT}_{\overbar{X}^{(i)}}(\overbar{f^{(i)}},\mu,\beta^{(i)})\pmod{p_i(\alpha)}.
    \end{equation}
    Therefore, applying these two equations iteratively yields 

    \begin{equation}\label{eq:lemma3.1}
    \begin{aligned}
        &\text{FFT}_{\overbar{X}}(\text{IFFT}_{\overbar{X}}(F_2,\mu,\omega_{2^\mu})+\ldots+\text{IFFT}_{\overbar{X}}(F_{2^{m-\mu}},\mu,\omega_{2^{m}-2^\mu}),\mu,0)\\ \equiv& \text{FFT}_{\overbar{X}^{(i)}}(\text{IFFT}_{\overbar{X}^{(i)}}(F_2^{(i)},\mu,\omega_{2^\mu}^{(i)})+\ldots+\text{IFFT}_{\overbar{X}^{(i)}}(F_{2^{m-\mu}}^{(i)},\mu,\omega_{2^{m}-2^\mu}^{(i)}),\mu,0)\pmod{p_i(\alpha)}.
    \end{aligned}
    \end{equation}

    For $i\in[t]$, since $F^{(i)}$ is a codeword of RS codes over the field $\mathbb{F}_2[x]/(p_i(x))$ and $F_2^{(i)},F_3^{(i)},\ldots,F_{2^{m-\mu}}^{(i)}$ are data symbols, then the following equation holds for $i\in[t]$ by \cite[Lemma 10]{2016FFTAlgorithm}.
    \begin{equation}\label{eq:encoding}
    \begin{aligned}
        F_1^{(i)} =& \text{FFT}_{\overbar{X}^{(i)}}(\text{IFFT}_{\overbar{X}^{(i)}}(F_2^{(i)},\mu,\omega_{2^\mu}^{(i)})+\ldots+\text{IFFT}_{\overbar{X}^{(i)}}(F_{2^{m-\mu}}^{(i)},\mu,\omega_{2^{m}-2^\mu}^{(i)}),\mu,0).
    \end{aligned}
    \end{equation}

    By Eq. \eqref{eq:lemma3.1} and \eqref{eq:encoding}, we have $$F_1\equiv F_1^{(i)}\equiv\text{FFT}_{\overbar{X}}(\text{IFFT}_{\overbar{X}}(F_2,\mu,\omega_{2^\mu})+\ldots+\text{IFFT}_{\overbar{X}}(F_{2^{m-\mu}},\mu,\omega_{2^{m}-2^\mu}),\mu,0)\pmod{p_i(\alpha)},$$ for $i\in[t]$. Therefore, the lemma is proved. 


    
\end{IEEEproof}

\subsection{Decoding Algorithm}

In \cite{2022NewEncoding}, there are two algorithms for decoding RS codes based on FFT and modular approach, i.e., the frequency-domain modular approach (FDMA) and the fast
modular approach (FMA). In this section, we will generalize the FDMA algorithm to our codes, and a similar 
generalization can be applied to the FMA algorithm.

Assume that the received vector over $\mathcal{R}_p$ is represented as 
\begin{align*}
    r = F + e = (f(\omega_0), f(\omega_1), \ldots, f(\omega_{2^m-1}))+(e_0, e_1, \ldots, e_{2^m-1}),
\end{align*}  where \( e=(e_0, e_1, \ldots, e_{2^m-1}) \) is the error pattern. Correspondingly, for $i=1,2,\ldots,t$, the received vector over the field $\mathbb{F}_2[x]/(p_i(x))$ is represented as \( r^{(i)} = F^{(i)} + e^{(i)} = (f^{(i)}(\omega_0^{(i)}), f^{(i)}(\omega_1^{(i)}), \ldots, f^{(i)}(\omega_{2^m-1}^{(i)})) + (e_0^{(i)}, e_1^{(i)}, \ldots, e_{2^m-1}^{(i)}) \), where \(e^{(i)}=(e_0^{(i)}, e_1^{(i)}, \ldots, e_{2^m-1}^{(i)})\) and $\Phi(e_j)=(e_j^{(1)},e_j^{(2)},\ldots,e_j^{(t)})$ for $j=0,1,\ldots,$\\$2^m-1$.

Since \(F^{(i)}=(f^{(i)}(\omega_0^{(i)}), f^{(i)}(\omega_1^{(i)}), \ldots, f^{(i)}(\omega_{2^m-1}^{(i)}))\) is the codeword of RS codes over \(\mathbb{F}_2[x]/(p_i(x))\) for $i=1,2,\ldots,t$, the trivial approach is to use the FDMA algorithm in \cite{2022NewEncoding} to decode $t$ RS codes to obtain the codewords \(\{F^{(i)}\}_{i\in[t]}\), then use \(\Phi^{-1}\) to obtain the original codeword \(F\). However, this method performs multiplication in the fields \(\{\mathbb{F}_2[x]/(p_i(x))\}_{i\in[t]}\), which has a higher multiplication complexity than in the cyclic polynomial ring. Our idea is to merge the decoding procedures over the \(t\) fields $\{\mathbb{F}_2[x]/(p_i(x))\}_{i\in[t]}$ and perform all multiplications in the cyclic polynomial ring without the need for modular decomposition.



We will present the algorithm in five steps:

\subsubsection{Syndrome}

For $i\in[t]$ and $l=0,1,\ldots,2^m-1$, define \begin{equation*}
    \begin{aligned}
        p_l &= s_0(v_0)^{l_0} s_1(v_1)^{l_1} \ldots s_{m-1}(v_{m-1})^{l_{m-1}}, \\
    p_l^{(i)} &= s_0^{(i)}(v_0^{(i)})^{l_0} s_1^{(i)}(v_1^{(i)})^{l_1} \ldots s_{m-1}^{(i)}(v_{m-1}^{(i)})^{l_{m-1}},
    \end{aligned}
\end{equation*} where \((l_0, l_1, \ldots, l_{m-1})\) is the binary representation of \(l\). 

Denote the coordinate vector of the syndrome polynomial \(u(x)\) with respect to $\overbar{X}$ as $\overbar{u}$, then combined with the following Lemma \ref{lemma:syndrome}, we show that we can compute $\overbar{u}$ by
\begin{equation*}
    \overbar{u}=\sum_{i=0}^{2^{m-\mu}-1} \text{IFFT}_{\overbar{X}}(r_i, \mu, \omega_{i \cdot 2^\mu}) / p_{2^m-2^\mu},
\end{equation*} where \(r_i = (r_{i \cdot 2^\mu}, r_{i \cdot 2^\mu + 1}, \ldots, r_{i \cdot 2^\mu + 2^\mu - 1})\) for $i=0,1,\ldots,2^{m-\mu}-1$. Subsequently, the syndrome\\\(\{u(\omega_i)\}_{i=0}^{2^\mu-1}\) is computed by $\text{FFT}_{\overbar{X}}(\overbar{u},\mu,0)$.

Next Lemma shows the relationship between the obtained syndromes and the corresponding syndromes of $t$ RS codes over fields.

\begin{lemma}\label{lemma:syndrome}
    For $i=0,1,\ldots,t$, denote \(u^{(i,field)}(x)\) as the syndrome polynomial of the RS code over the field \(\mathbb{F}_2[x]/(p_i(x))\) whose coordinate vector with respect to $\overbar{X}^{(i)}$ is denoted as $\overbar{u^{(i,field)}}$, then
    \begin{equation*}
    u(\omega_j)\equiv u^{(i,field)}(\omega_j^{(i)})\pmod{p_i(\alpha)},j=0,1,\ldots,2^\mu-1.
\end{equation*} 
\end{lemma}

\begin{IEEEproof}
     By \cite[Eq.9]{2022NewEncoding}, for $i=1,2,\ldots,t$, \(\overbar{u^{(i,field)}}\) can be computed by \begin{align*}
        \overbar{u^{(i,field)}}=\sum_{j=0}^{2^{m-\mu}-1} \text{IFFT}_{\overbar{X}^{(i)}}(r_j^{(i)}, \mu, \omega^{(i)}_{j \cdot 2^\mu}) / p^{(i)}_{2^m-2^\mu},
     \end{align*} where $r_j^{(i)}=(r_{j \cdot 2^\mu}^{(i)}, r_{j \cdot 2^\mu + 1}^{(i)}, \ldots, r_{j \cdot 2^\mu + 2^\mu - 1}^{(i)})$ for $i=1,2,\ldots,t$ and $j=0,1,\ldots,2^{m-\mu}-1$.
    For $i=1,2,\ldots,t$, applying Eq. \eqref{eq:FFTlemma} and \eqref{eq:IFFTlemma} to the definition of $\overbar{u}$ iteratively, we obtain
\begin{equation*}
    \begin{aligned}
        \overbar{u}\equiv  \sum_{j=0}^{2^{m-\mu}-1} \text{IFFT}_{\overbar{X}^{(i)}}(r_j^{(i)}, \mu, \omega^{(i)}_{j \cdot 2^\mu}) / p^{(i)}_{2^m-2^\mu}
        \equiv  \overbar{u^{(i,field)}}\pmod{p_i(\alpha)},
    \end{aligned}
\end{equation*}
    Therefore, $u(x)\equiv u^{(i,field)}(x^{(i)})\pmod{p_i(\alpha)}$ for $i=1,2,\ldots,t$ and the lemma is proved.
\end{IEEEproof}

\subsubsection{Solving the Key Equation by FDMA}

Given the syndrome $\{u(\omega_i)\}_{i=0,1,\ldots,2^\mu-1}$, the FDMA algorithm outputs $(\lambda(x),z(x))$, where $\lambda(x)$ is the error locator polynomial and $z(x)$ is the error evaluation polynomial.

The FDMA algorithm corresponding to our codes is presented in Algorithm \ref{alg:FDMA}, in which $\mathbb{I}$ is the indicator function (i.e., $\mathbb{I}_A(x)=\begin{cases}
x, & \text{ if } A\  \text{is true;} \\ 
0, & \text{ if } A\  \text{is false.} 
\end{cases}$) and the extended IFFT algorithm is identical to \cite[Algorithm 5]{2022NewEncoding}. In Algorithm \ref{alg:FDMA}, for $i=1,2,\ldots,t$ and \(r_i \in \mathbb{F}_2[x]/(p_i(x))\), $\Phi^{-1}(r_1,\ldots,r_t)$ can be computed as follows by the inverse mapping of the Chinese remainder theorem:
\[
\Phi^{-1}(r_1,\ldots,r_t) = \sum_{i=1}^t r_i \cdot \frac{M_p(\alpha)}{p_i(\alpha)} \left[\left(\frac{M_p(\alpha)}{p_i(\alpha)}\right)^{-1}\right]_{p_i},
\]
where $\left[\left(\frac{M_p(\alpha)}{p_i(\alpha)}\right)^{-1}\right]_{p_i}$ represents the inverse of $\frac{M_p(\alpha)}{p_i(\alpha)}$ in $\mathbb{F}_2[\alpha]/(p_i(\alpha))$. Since $\frac{M_p(\alpha)}{p_i(\alpha)}\left[\left(\frac{M_p(\alpha)}{p_i(\alpha)}\right)^{-1}\right]_{p_i}$ can be computed in advance, computing $\Phi^{-1}$ requires $t$ multiplications in $\mathcal{R}_p$. The correctness of the extended IFFT algorithm used in Algorithm \ref{alg:FDMA} can be proved similarly to the FFT algorithm.

The next Lemma \ref{lemma:FDMA} shows the relationship between the obtained error locator polynomial and the error evaluation polynomial to the corresponding two polynomials of $t$ RS codes over fields, which shows the rationality of our FDMA algorithm.

\begin{algorithm}
\caption{$\text{FDMA}$ over $\mathcal{R}_p$}\label{alg:FDMA}
\begin{algorithmic}[1]
\Require $\{\omega_i,u(\omega_i)\},i=0,1,\ldots,2^\mu-1$.
\Ensure $(\lambda(x),z(x))$ that are represented with respect to $\overbar{X}$ and rank[$\lambda^{(i)}(x),z^{(i)}(x)$], $i\in[t]$.
\State \textbf{Initialize:} $d_i^0=u(\omega_{i-1})$, $g_i^0=1$, for $i=1,2,\ldots,2^\mu$, $W(\omega_i)=0$, $V(\omega_i)=0$, $i=0,1,\ldots,2^{\mu-1}$, $r^0_{1,(k)}=0$, $r^0_{2,(k)}=1$, $k\in[t]$.
\For{$j=1,2,\ldots,2^\mu$}
    \For{$k=1,2,\ldots,t$}
        \If{$g_j^{j-1}\equiv0\pmod{p_k(\alpha)}$ or ($d_j^{j-1}\equiv0\pmod{p_k(\alpha)}$ and $r_{1,(k)}^{j-1}<r_{2,(k)}^{j-1}$)}
            \State $\text{Flag}^{(k)}=1$.
            \State $r^j_{1,(k)}=r^{j-1}_{2,(k)}$, $r^j_{2,(k)}=r^{j-1}_{1,(k)}+2$
        \Else
            \State $\text{Flag}^{(k)}=2$.
            \State $r^j_{1,(k)}=r^{j-1}_{1,(k)}$, $r^j_{2,(k)}=r^{j-1}_{2,(k)}+2$
        \EndIf
        \State  $\psi_{21}=\Phi^{-1}(\mathbb{I}_{\text{Flag}^{(1)} = 1}(\omega_i^{(1)}-\omega_{j-1}^{(1)}), \mathbb{I}_{\text{Flag}^{(2)} = 1}(\omega_i^{(2)}-\omega_{j-1}^{(2)}),\ldots,\mathbb{I}_{\text{Flag}^{(t)} = 1}(\omega_i^{(t)}-\omega_{j-1}^{(t)}))$
        \State $\psi_{22}=\Phi^{-1}(\mathbb{I}_{\text{Flag}^{(1)} = 2}(\omega_i^{(1)}-\omega_{j-1}^{(1)}), \mathbb{I}_{\text{Flag}^{(2)} = 2}(\omega_i^{(2)}-\omega_{j-1}^{(2)}),\ldots,\mathbb{I}_{\text{Flag}^{(t)} = 2}(\omega_i^{(t)}-\omega_{j-1}^{(t)}))$
        \State $\Psi_j(\omega_i)$=$\begin{pmatrix}
            -g_j^{j-1}&d_j^{j-1}\\
            \psi_{21} & \psi_{22}
        \end{pmatrix}$, $i=0,1,..,2^\mu-1$
    \EndFor
    \For{$i=j+1,j+2,\ldots,2^\mu$}
        \State $\begin{pmatrix}
            d_i^j\\
            g_i^j
        \end{pmatrix}$ = $\Psi_j(\omega_{i-1})$$\begin{pmatrix}
            d_i^{j-1}\\
            g_i^{j-1}
        \end{pmatrix}$
    \EndFor
    \For{$i=0,1,\ldots,2^{\mu-1}$}
        \State $\begin{pmatrix}
                W(\omega_i)\\
                V(\omega_i)
            \end{pmatrix}$ = $\Psi_j(\omega_{i})$$\begin{pmatrix}
                W(\omega_i)\\
                V(\omega_i)
            \end{pmatrix}$
    \EndFor
\EndFor
\For{$k=1,2,\ldots,t$}
    \If{$r_{1,(k)}^{2^\mu}$ > $r_{2,(k)}^{2^\mu}$}
        \State $\lambda^{(k)}(\omega_i)=V(\omega_i)$, $i=0,1,\ldots,2^{\mu-1}$.
    \Else
        \State $\lambda^{(k)}(\omega_i)=W(\omega_i)$, $i=0,1,\ldots,2^{\mu-1}$.
    \EndIf
\EndFor
\State $\lambda(\omega_i)=\Phi^{-1}(\lambda^{(1)}(\omega_i),\lambda^{(2)}(\omega_i),\ldots,\lambda^{(t)}(\omega_i))$, $i=0,1,\ldots,2^{\mu-1}$.
\State $z(\omega_i)=\lambda(\omega_i)u(\omega_i)$, $i=0,1,\ldots,2^{\mu-1}$.
\State $\lambda(x)=\text{extended IFFT}_{\overbar{X}}((\lambda(\omega_0),\ldots,\lambda(\omega_{2^{\mu-1}})),\mu-1,0)$
\State $z(x)=\text{extended IFFT}_{\overbar{X}}((z(\omega_0),\ldots,z(\omega_{2^{\mu-1}})),\mu-1,0)$
\State \Return $(\lambda(x),z(x))$ and rank[$\lambda^{(i)}(x),z^{(i)}(x)$]=min($r_{1,(i)}^{2^\mu}$,$r_{2,(i)}^{2^\mu}$), $i\in[t]$.
\end{algorithmic}
\end{algorithm}

\begin{lemma}\label{lemma:FDMA}
    For $i=1,2,\ldots,t$, denote the error locator polynomial of RS codes over \(\mathbb{F}_2[x]/(p_i(x))\) as \(\lambda^{(i,field)}(x)\) and denote the corresponding error evaluation polynomial as \(z^{(i,field)}(x)\), then $(\lambda(x),z(x))$ obtained by our FDMA algorithm in Algorithm~\ref{alg:FDMA} satisfies:
    \begin{equation*}
        \begin{aligned}
            \lambda(x)&\equiv\lambda^{(i,field)}(x^{(i)})\pmod{p_i(\alpha)},\\
            z(x)&\equiv z^{(i,field)}(x^{(i)})\pmod{p_i(\alpha)}.
        \end{aligned}
    \end{equation*}
    
\end{lemma}

\begin{IEEEproof}
    For $i=1,2,\ldots,t$, if all elements in the computation process are taken modulo $p_i(\alpha)$ component-wise, the input is the same as the input for the FDMA algorithm over $\mathbb{F}_2[x]/(p_i(x))$ as described in \cite[Algorithm 4]{2022NewEncoding} due to Lemma \ref{lemma:syndrome}. Furthermore, it can be verified that steps 1-30 are also the same as in the FDMA over $\mathbb{F}_2[x]/(p_i(x))$ when taken modulo \(p_i(\alpha)\). Therefore, we obtain
    \begin{equation}\label{eq:FDMAlemma1}
        \begin{aligned}
            \lambda(\omega_j)&\equiv\lambda^{(i,field)}(\omega_j^{(i)})\pmod{p_i(\alpha)},\\
        \end{aligned}
    \end{equation}
    where $j=0,1,\ldots,2^{\mu-1}$.

    Similar to Eq. \eqref{eq:FFTlemma}, we can obtain the following property of extended IFFT. For any $f \in \mathcal{R}_p[x]$ and $i\in[t]$, the following holds:
    \begin{equation}\label{eq:FDMAlemma2}
        \begin{aligned}
            \text{extended IFFT}_{\overbar{X}}(\overbar{f},\mu,\beta)
            \equiv \text{extended IFFT}_{\overbar{X}^{(i)}}(\overbar{f^{(i)}},\mu,\beta^{(i)})\pmod{p_i(\alpha)}.
        \end{aligned}
    \end{equation} 
    
    Therefore, combining Eq. \eqref{eq:FDMAlemma1} and Eq. \eqref{eq:FDMAlemma2}, we have:\begin{equation*}
        \begin{aligned}
            \lambda(x)=&\text{extended IFFT}_{\overbar{X}}((\lambda(\omega_0),\ldots,\lambda(\omega_{2^{\mu-1}})),\mu-1,0)\\
            \equiv&\text{extended IFFT}_{\overbar{X}^{(i)}}((\lambda^{(i,field)}(\omega_0^{(i)}),\ldots,\lambda^{(i,field)}(\omega^{(i)}_{2^{\mu-1}})),\mu-1,0)\\
            \equiv& \lambda^{(i,field)}(x^{(i)})\pmod{p_i(\alpha)},
        \end{aligned}
    \end{equation*} where the last equation is obtained by Step 23 in \cite[Algorithm 4]{2022NewEncoding}.
    The equation for \(z(x)\) can be similarly proved.

\end{IEEEproof}

\subsubsection{Chein search}
The Chein search algorithm to find the error locations corresponding to our codes is given by Algorithm~\ref{alg:CheinSearch}.
Assuming the set of error locations for RS codes over the field $\mathbb{F}_2[x]/(p_i(x))$ is denoted as $E^{(i,true)}$, and the set of error locations obtained by our Chein search algorithm is denoted as $E^{(i)}$ for $i\in[t]$. 

\begin{algorithm}
\caption{Chein search algorithm over $\mathcal{R}_p$}\label{alg:CheinSearch}
\begin{algorithmic}[1]
\Require $\overbar{\lambda}=(\overbar{\lambda}_0,\overbar{\lambda}_1,\ldots,\overbar{\lambda}_{2^\mu-1}) $
\Ensure the set of error locations $E^{(i)}$ for $i\in[t]$
\For{$l=0,1,\ldots,2^{m-\mu}-1$}
    \State $(\lambda(\omega_{l\cdot 2^\mu}),\ldots,\lambda(\omega_{l\cdot 2^\mu+2^\mu-1}))$=$\text{FFT}_{\overbar{X}}(\overbar{\lambda},\mu,\omega_{l\cdot 2^\mu})$
    \For{$i=1,2,\ldots,t$}
        \For{$j=l\cdot 2^\mu,\ldots,l\cdot 2^\mu+2^\mu-1$}
            \If{$\lambda(\omega_j)\equiv0\pmod{p_i(\alpha)}$}
                \State Add $\omega_j$ to the set $E^{(i)}$
            \EndIf
        \EndFor
    \EndFor

\EndFor
\For{$i=1,2,\ldots,t$}
    \State $\lambda^{(i)}(x)=\lambda(x)\pmod{p_i(\alpha)}$
    \If{$\text{deg}(\lambda^{(i)}(x))$ > $|E^{(i)}|$}
        \State Uncorrectable error is detected, and the decoding procedure terminates.
    \EndIf
\EndFor
\end{algorithmic}
\end{algorithm}

The next Lemma shows that $E^{(i,true)} = E^{(i)}$ for $i\in[t]$, which shows the rationality of our Chein search algorithm.

\begin{lemma}\label{lemma:Chein Search}
    $E^{(i,true)} = E^{(i)}$ for $i\in[t]$.
\end{lemma}

\begin{IEEEproof}
    By Lemma \ref{lemma:FDMA}, it can be deduced that $\lambda(\omega_j) \equiv \lambda^{(i,field)}(\omega_j^{(i)})\pmod{p_i(\alpha)} $ for $i=1,2,\ldots,t$ and $j=0,1,\ldots,2^m-1$. Therefore, Algorithm \ref{alg:CheinSearch} indeed computes the roots of $\lambda^{(i,field)}$ on $\mathbb{F}_2[x]/(p_i(x))=\{\omega_j^{(i)}\}_{j=0,1,\ldots,2^m-1}$ for $i\in[t]$, which are exactly the error locations for RS codes over $\mathbb{F}_2[x]/(p_i(x))$. This establishes the proof that $E^{(i,true)} = E^{(i)}$ for $i\in[t]$.
\end{IEEEproof}

\subsubsection{Formal Derivative}

The algorithm for formal derivative is identical to that in \cite{2016FFTAlgorithm}, which computes the derivative $\lambda'(x)$ of the error locator polynomial $\lambda(x)$.

\subsubsection{Forney's Formula}

We present Forney's formula for our codes to retrieve the original data symbols $\{g_l\}_{l=0}^{k-1}=\{f(\omega_l)\}_{l=2^\mu}^{2^m-1}$ as follows.

For $l=2^\mu,2^\mu+1,\ldots,2^m-1$, if $\omega_l\notin \cup_{i}E^{(i)}$, it implies that there is no error at $\omega_l$. Therefore, let $f(\omega_l) = r(\omega_l)$. Otherwise, calculate $h(\omega_l) = \frac{z(\omega_l)}{\lambda'(\omega_l)s_\mu(\omega_l)}$ and $e(\omega_l) = \Phi^{-1}(\mathbb{I}_{l\in E^{(1)}}h(\omega_l),\ldots,\mathbb{I}_{l\in E^{(t)}}h(\omega_l))$, then let $f(\omega_l) = r(\omega_l) + e(\omega_l)$. The next theorem shows that the obtained $\{f(\omega_l)\}_{l=2^\mu}^{2^m-1}$ are indeed the original data symbols.

\begin{theorem}\label{theorem:the7}
If the number of errors is no larger than the error correction capability, then our decoding algorithm can retrieve the original data symbols.
\end{theorem}

\begin{IEEEproof}
    It can be verified by Lemma \ref{lemma:FDMA} that the steps of Forney's formula for our codes are equivalent to the following steps when taken modulo $p_i(\alpha)$ for $i=1,2,\ldots,t$: For $l=2^\mu,2^\mu+1,\ldots,2^m-1$, if $\omega_l \in E^{(i)}$, then let $f(\omega_l) = r^{(i)}(\omega_l^{(i)}) + \frac{z^{(i)}(\omega_l^{(i)})}{\lambda^{(i)\prime}(\omega_l^{(i)})s^{(i)}_\mu(\omega_l^{(i)})}$, otherwise let $f(\omega_l)=r^{(i)}(\omega_l^{(i)})$. By Lemma \ref{lemma:Chein Search}, the steps are indeed the procedure of Forney's formula over the field $\mathbb{F}_2[x]/(p_i(x))$ presented in \cite{2022NewEncoding}. Therefore for $i=1,2,\ldots,t$ and $l=2^\mu,2^\mu+1,\ldots,2^m-1$, $f(\omega_l)$ is the original data symbol of RS codes over $\mathbb{F}_2[x]/(p_i(x))$ when taken modulo $p_i(\alpha)$, i.e., $f(\omega_l)\equiv f^{(i)}(\omega_l^{(i)})\pmod{p_i(\alpha)}$, which implies that $\{f(\omega_l)\}_{l=2^\mu}^{2^m-1}$ obtained by Forney's formula are indeed the original data symbols of our codes.
\end{IEEEproof}


We can show that the asymptotic number of addition and multiplication operations involved in the encoding/decoding procedure of our codes is identical to \cite{2022NewEncoding}.

\section{Reduce Encoding/decoding complexity}\label{sec:ReduceXOR}

In this section, we propose several methods to reduce encoding/decoding complexity by minimizing the XOR operations required for multiplication in the cyclic polynomial ring $\mathcal{R}_p$.

\subsection{Efficient Multiplication by Circular Shifts}

Due to the equation $1+x^p = (1+x)M_p(x)$ in $\mathbb{F}_2[x]$, all multiplication in the ring $\mathcal{R}_p$ during encoding/decoding procedure can be replaced with the multiplication in $\mathbb{F}_2[x]/(1+x^p)$. It is sufficient to take the final results of the encoding/decoding procedure modulo $M_p(x)$ to ensure the correctness of the outcomes. The multiplication in $\mathbb{F}_2[x]/(1+x^p)$ can be efficiently performed using circular shifts, as outlined in Algorithm \ref{alg:Multiplication}, where $[a]_p$ represents the integer $b\in\{0,1,\ldots,p-1\}$ such that $a\equiv b\pmod{p}$.

\begin{algorithm}
\caption{Multiplication in $\mathbb{F}_2[x]/(1+x^p)$}\label{alg:Multiplication}
\begin{algorithmic}[1]
\Require $f(\alpha)=\sum\limits_{i=0}^{p-1}f_i\alpha^i,g(\alpha)=\sum\limits_{i=0}^{p-1}g_i\alpha^i$
\Ensure $h(\alpha)=\sum\limits_{i=0}^{p-1}h_i\alpha^i=f(\alpha)g(\alpha)$  
\For{$i=0,1,\ldots,p-1$}
    \If{$f_i$ = $1$}
        \For{$j=0,1,\ldots,p-1$}
            \State $h_{[i+j]_p}=h_{[i+j]_p}+g_j$
        \EndFor
    \EndIf
\EndFor
\end{algorithmic}
\end{algorithm}

In comparison to the field multiplication where additional modulo operations are required after polynomial multiplication, multiplication in the ring $\mathbb{F}_2[x]/(1+x^p)$ involves fewer XOR operations.

\subsection{Preprocessing in the Ring $\mathbb{F}_2[x]/(1+x^p)$}\label{subsec:Preprocess}

According to Algorithm \ref{alg:Multiplication}, the more zeros in the coefficients of $f(\alpha)\in \mathbb{F}_2[x]/(1+x^p) $, the fewer XOR operations are needed. Therefore, we present the following preprocessing steps: Iterate through the $p$ coefficients of $f(\alpha)$ and count the number of zeros, denoted as $n$. If $n < \frac{p}{2}$, then let $f_1(\alpha) = f(\alpha) + M_p(\alpha)$ and replace $f$ with $f_1$, resulting in $p - n$ zeros in the coefficients of $f_1(\alpha)$, which is more than before preprocessing. Since $f(\alpha)g(\alpha) \equiv f_1(\alpha)g(\alpha) \pmod{M_p(\alpha)}$ holds, the preprocessing does not affect the correctness.

We categorize all multiplications into two cases. In the first case, both factors are dependent on the received codeword and such multiplications are only computed during the FDMA algorithm when calculating $g_{j-1}^j d^{j-1}_i$ and $d_{j-1}^j g^{j-1}_i$. For this case, it is sufficient to preprocess $g_{j-1}^j$ and $d_{j-1}^j$ for $j=1,2,\ldots,2^\mu$. The XOR operations involved in preprocessing are counted in the decoding procedure. In the second case, at least one of the factors is independent of the received codeword and can be computed in advance. For example, when performing FFT algorithm in Algorithm \ref{alg:FFT}, the factor $\frac{s_{\tau-1}(\beta)}{s_{\tau-1}(v_{\tau-1})}$ can be computed in advance. Therefore, we preprocess the factor $\frac{s_{\tau-1}(\beta)}{s_{\tau-1}(v_{\tau-1})}$ and the additional XOR operations required are not counted in the encoding/decoding procedure.



\subsection{Selection of $v_j^{(i)}$}\label{subsec:SelectionOfvi}

Since the selection of $v_j^{(i)}$ for $j=0,1,\ldots,m-1$ and $i=1,2,\ldots,t$ only needs to satisfy the conditions given in \ref{subsec:Construction}, and throughout the encoding/decoding procedure, 
the majority of multiplications stem from FFT and IFFT, with nearly identical parameters. Therefore, a meticulous choice of $v_j^{(i)}$ for $j=0,1,\ldots,m-1$ and $i=1,2,\ldots,t$ will be introduced to reduce the number of XORs required for the multiplication in the FFT and IFFT algorithms.

Taking the computation of the syndrome as an example, it involves calculating $\text{IFFT}_{\overbar{X}}(r_i, \mu, \omega_{i\cdot 2^\mu})$ for $i=0,1,\ldots,2^{m-\mu}-1$. For each $i$, there are a total of $\frac{1}{2}\mu 2^\mu$ multiplications in $\text{IFFT}_{\overbar{X}}(r_i, \mu, \omega_{i\cdot 2^\mu})$. Each multiplication involves a factor of the form $\frac{s_{\tau-1}(\beta)}{s_{\tau-1}(v_{\tau-1})}$, where $\tau=1,\ldots,\mu$. Assuming $c_\beta$ represents the sum of the number of non-zero coefficients of $\frac{s_{\tau-1}(\beta)}{s_{\tau-1}(v_{\tau-1})}$ after preprocessing in all multiplications during the calculation of $\text{IFFT}_{\overbar{X}}(r_0, \mu, \beta)$, where the first parameter $r_0$ in IFFT is independent of $c_\beta$. The objective is to minimize $\sum\limits_{i=0}^{2^{m-\mu}-1} c_{\omega_{i\cdot 2^\mu}}$. The variables are the bases $\{v_j^{(i)}\}_{j=0,1,\ldots,m-1}$, where $i=1,2,\ldots,t$. 




When the parameters $m$ and $\mu$ of the code are small, an exhaustive algorithm can be employed. Otherwise, we start by randomly choosing $\{v_j^{(i)}\}_{j=0,1,\ldots,\mu-1}$ for $i=1,2,\ldots,t$ to determine the expression of $s_{\mu-1}(x)$ and the value of $v_{\mu-1}$. The subsequent bases for $j=\mu,\ldots,m-1$ can be determined by the following approximate algorithm. Let $r_0=\operatorname*{argmin}_{r\in\mathcal{R}_p} c_r$. If $r_0^{(i)}$ is linearly independent of $\{v_j^{(i)},j=0,1,\ldots,\mu-1\}$ for all $i\in[t]$, then set $v_\mu^{(i)}=r_0^{(i)}$ for $i=1,2,\ldots,t$. Otherwise, select the $r$ that has the second smallest $c_r$ and check if it satisfies the condition. Continue this process until all bases are determined.

\section{Comparison of encoding/decoding complexity}\label{sec:Comparison}

In this section, we compare the encoding/decoding complexity of our codes with RS codes over finite fields presented in \cite{2022NewEncoding}. First, we discuss the parameters. 

$\mathcal{R}_p$ is a field if and only if $2$ is a primitive element in $\mathbb{F}_p$ \cite{1993NewArrayCodes}, which occurs when $p=3,5,11,13,19,\cdots$. In this case, we can directly apply the encoding/decoding algorithms for RS codes over fields in \cite{2022NewEncoding} and reduce encoding/decoding complexity by the methods discussed in Section \ref{sec:ReduceXOR}. Now, we consider the case when $\mathcal{R}_p$ is not a field. The code lengths of our codes for different $p$ are shown in Table \ref{tab:CodeLength}.



 \begin{table}[htbp]
   \renewcommand{\arraystretch}{1.3}
   \caption{The code lengths of our codes for different $p$.}
   \label{tab:CodeLength}
   \centering
    \resizebox{.5\linewidth}{!}{\begin{tabular}{|c|c|c|c|c|c|c|c|}
			\hline	$p$&7&9&15&17&21&23&25\\
			\hline
			Code Length&$2^3$&$2^2$&$2^2$&$2^8$&$2^2$&$2^{11}$&$2^4$\\
			\hline
		\end{tabular}}
 \end{table}



Due to the code length being much smaller than the size of the ring $\mathcal{R}_p$ when the irreducible factorization of $M_p(x)=\prod\limits_{i=1}^t p_i(x)$ has a large $t$, we find that there is no benefit in using our codes in such cases through numerical experiments. Therefore, the subsequent discussion focuses the case of $t=2$ and $\deg(p_1(x))=\deg(p_2(x))$, corresponding to $p=7,17,23,\cdots$.

We compare the encoding/decoding complexity for our codes with RS codes over finite fields presented in \cite{2022NewEncoding}. Considering the parameters of our codes: Since $|\mathcal{R}_p|=2^{p-1}$, each data symbol has $p-1$ bits, and the code length is $2^{\frac{p-1}{2}}$. For the comparison, RS codes over finite fields have two options. The first option is to take RS codes defined over $\mathbb{F}_{2^{p-1}}$, satisfying $|\mathbb{F}_{2^{p-1}}|=|\mathcal{R}_p|$, and the code length can reach $2^{\frac{p-1}{2}}$. The second option is to divide each data symbol into two parts, each having $\frac{p-1}{2}$ bits. This enables encoding/decoding using two RS codes over \(\mathbb{F}_{2^{\frac{p-1}{2}}}\) with the resulting data symbols, and each RS code can achieve a code length of $2^{\frac{p-1}{2}}$. We choose the second approach for comparison because it possesses lower encoding/decoding complexity.

For the parameters \(p=23\), \(m=11\), and \(\mu=6\), we consider our (2048, 1984) RS code over \(\mathcal{R}_{23} = \mathbb{F}_2[x]/(1+x+\ldots+x^{22})\). The encoding and decoding procedures are performed as described in Section \ref{sec:encoding} and \ref{sec:ReduceXOR}. For comparison, two (2048, 1984) RS codes over \(\mathbb{F}_2[x]/(1+x^2+x^{11})\) are selected, using the encoding/decoding algorithms described in \cite{2022NewEncoding}. The choice of the irreducible polynomial \(1+x^2+x^{11}\) is made to reduce the number of XORs in field multiplication. The comparisons are carried out by repeatedly performing encoding and decoding and counting the average number of XORs at each step. Table \ref{tab:XORcomparison} shows the detailed comparisons.
\begin{small}
\begin{table}[htbp]
   \renewcommand{\arraystretch}{1.3}
   \caption{Comparison of the number of XOR Operations during the Encoding/Decoding Procedure of our (2048, 1984) code over the Ring \(\mathbb{F}_2[x]/(1+x+\ldots+x^{22})\) and Two (2048, 1984) RS Codes over the Field \(\mathbb{F}_2[x]/(1+x^2+x^{11})\).
}
   \label{tab:XORcomparison}
   \centering
    \resizebox{.5\linewidth}{!}{\begin{tabular}{ |c|c|c| }
 \hline
 \multirow{2}{*}{Steps} & \multicolumn{2}{|c|}{Average XOR Operations} \\
 \cline{2-3} 
 & RS codes over $\mathcal{R}_{23}$ & Two RS codes over field \\
 \hline
 Encoding & 1389200.4 & 1692843.6 \\
 \hline
 Syndrome & 1420710 & 1736304 \\
 \hline
 Key equation & 2081194.8 & 2210871 \\
 \hline
 Chein search & 1489731.4 & 1482622.4 \\
 \hline
 Formal derivative & 44896 & 30605.6 \\
 \hline
 Forney's formula & 242024.5 & 250949.2 \\
 \hline
 Total decoding & 5278556.7 & 5711352.2 \\
 \hline
\end{tabular}}
 \end{table}\end{small}

In the encoding/decoding procedure of our codes, we find a choice of $\{v_i\}_{i=0,1,\ldots,m-1}$ according to the method in \ref{subsec:SelectionOfvi} as follows: \(v_0 = \alpha^6\), \(v_1 = \alpha^5\), \(v_2 = \alpha^4\), \(v_3 = \alpha^3\), \(v_4 = \alpha^7\), \(v_5 = \alpha^2\), \(v_6 = \alpha^8 + \alpha^3\), \(v_7 = \alpha\), \(v_8 = \alpha^9 + \alpha^8 + \alpha^3 + \alpha + 1\), \(v_9 = \alpha^{10} + \alpha^9 + \alpha^7 + \alpha^3 + \alpha^2 + \alpha\), \(v_{10} = \alpha^9 + \alpha^7 + \alpha^3\), where $\alpha$ is the indeterminate of $\mathcal{R}_{23}$.

According to Table \ref{tab:XORcomparison}, the proposed algorithm for our codes over the cyclic polynomial ring reduces the number of XORs by 17.9\% during encoding and 7.5\% during decoding.

\section{Conclusion}\label{sec:Conclusion}

In this paper, we construct RS codes over the cyclic polynomial ring $\mathcal{R}_p$. In addition, we generalize the encoding/decoding algorithms for RS codes over finite fields in \cite{2022NewEncoding} to our codes. We then reduce encoding/decoding complexity by the structure of the cyclic polynomial ring $\mathcal{R}_p$. Finally, we demonstrate that our codes can reduce 17.9\% encoding complexity and 7.5\% decoding complexity compared with the RS codes over finite fields when $(n,k)=(2048,1984)$.

\ifCLASSOPTIONcaptionsoff
\newpage
\fi

\bibliographystyle{IEEEtran}
\bibliography{ref}

\begin{thebibliography}{10}
\providecommand{\url}[1]{#1}
\csname url@samestyle\endcsname
\providecommand{\newblock}{\relax}
\providecommand{\bibinfo}[2]{#2}
\providecommand{\BIBentrySTDinterwordspacing}{\spaceskip=0pt\relax}
\providecommand{\BIBentryALTinterwordstretchfactor}{4}
\providecommand{\BIBentryALTinterwordspacing}{\spaceskip=\fontdimen2\font plus
\BIBentryALTinterwordstretchfactor\fontdimen3\font minus \fontdimen4\font\relax}
\providecommand{\BIBforeignlanguage}[2]{{%
\expandafter\ifx\csname l@#1\endcsname\relax
\typeout{** WARNING: IEEEtran.bst: No hyphenation pattern has been}%
\typeout{** loaded for the language `#1'. Using the pattern for}%
\typeout{** the default language instead.}%
\else
\language=\csname l@#1\endcsname
\fi
#2}}
\providecommand{\BIBdecl}{\relax}
\BIBdecl

\bibitem{1960ReedPolynomial}
I.~S. Reed and G.~Solomon, ``{Polynomial Codes over Certain Finite Fields},'' \emph{J. Soc. Ind. Appl. Math.}, vol.~8, no.~2, pp. 300--304, 1960.

\bibitem{2015FasterAlgorithms}
M.~F.~I. Chowdhury, C.-P. Jeannerod, V.~Neiger, {\'E}.~Schost, and G.~Villard, ``{Faster Algorithms for Multivariate Interpolation With Multiplicities and Simultaneous Polynomial Approximations},'' \emph{IEEE Trans. Inf. Theory}, vol.~61, no.~5, pp. 2370--2387, 2015.

\bibitem{2020FastEncoding}
N.~Tang and Y.~Lin, ``{Fast Encoding and Decoding Algorithms for Arbitrary $(n,k)$ Reed-Solomon Codes Over $\mathbb{F}_{2^m}$ },'' \emph{IEEE Communi. Lett.}, vol.~24, no.~4, pp. 716--719, 2020.

\bibitem{2022NewEncoding}
N.~Tang and Y.~S. Han, ``{A New Decoding Method for Reed–Solomon Codes Based on FFT and Modular Approach},'' \emph{IEEE Trans. Commun.}, vol.~70, no.~12, pp. 7790--7801, 2022.

\bibitem{2023EfficientInterpolation}
W.~K. Kadir, H.-Y. Lin, and E.~Rosnes, ``{Efficient Interpolation-Based Decoding of Reed-Solomon Codes},'' in \emph{2023 IEEE Int. Symp. Inf. Theory (ISIT)}, 2023, pp. 997--1002.

\bibitem{1993NewArrayCodes}
M.~Blaum and R.~Roth, ``{New array codes for multiple phased burst correction},'' \emph{IEEE Trans. Inf. Theory}, vol.~39, no.~1, pp. 66--77, 1993.

\bibitem{hou2016}
H.~{Hou}, K.~W. {Shum}., M.~Chen, and H.~{Li}, ``{BASIC Codes: Low-Complexity Regenerating Codes for Distributed Storage Systems},'' \emph{IEEE Trans. Inf. Theory}, vol.~62, no.~6, pp. 3053--3069, 2016.

\bibitem{hou2018a}
H.~{Hou} and Y.~S. {Han}, ``{A New Construction and an Efficient Decoding Method for Rabin-Like Codes},'' \emph{IEEE Trans. Commun.}, vol.~66, no.~2, pp. 521--533, 2018.

\bibitem{Hou2018form}
H.~{Hou}, Y.~S. {Han}, K.~W. {Shum}, and H.~{Li}, ``{A Unified Form of EVENODD and RDP Codes and Their Efficient Decoding},'' \emph{IEEE Trans. Commun.}, vol.~66, no.~11, pp. 5053--5066, 2018.

\bibitem{hou2021generalization}
H.~Hou, Y.~S. Han, P.~P. Lee, Y.~Wu, G.~Han, and M.~Blaum, ``{A Generalization of Array Codes with Local Properties and Efficient Encoding/Decoding},'' \emph{accepted in IEEE Trans. Information Theory, (early access: https://ieeexplore.ieee.org/document/9868795)}, 2022.

\bibitem{HOU2019}
H.~{Hou}, Y.~S. {Han}, P.~P.~C. {Lee}, Y.~{Hu}, and H.~{Li}, ``{A New Design of Binary MDS Array Codes with Asymptotically Weak-Optimal Repair},'' \emph{IEEE Trans. Inf. Theory}, vol.~65, no.~11, pp. 7095–--7113, 2019.

\bibitem{2010SubspacePoly}
S.~Gao and T.~Mateer, ``{Additive Fast Fourier Transforms Over Finite Fields},'' \emph{IEEE Trans. Inf. Theory}, vol.~56, no.~12, pp. 6265--6272, 2010.

\bibitem{2016FFTAlgorithm}
S.-J. Lin, T.~Y. Al-Naffouri, and Y.~S. Han, ``{FFT Algorithm for Binary Extension Finite Fields and Its Application to Reed–Solomon Codes},'' \emph{IEEE Trans. Inf. Theory}, vol.~62, no.~10, pp. 5343--5358, 2016.

\end{thebibliography}
\clearpage

\appendices

\end{document}